\documentclass[%
 aip,
cp,  
 amsmath,amssymb,
 reprint,%
]{revtex4-2}

\usepackage{graphicx}
\usepackage{dcolumn}
\usepackage{bm}

\usepackage[utf8]{inputenc}
\usepackage[T1]{fontenc}
\usepackage{mathptmx} 

\begin{document}

\title{A new scheme of negative muon beam generation with MuCF}

\author{Yoshiharu Mori} 
 \email[Corresponding author: ]{mori.yoshiharu.4w@kyoto-u.ac.jp}
\affiliation{
 Institute for Integrated Radiation and Nuclear Science, Kyoto University, Kumatori Osaka 590-0494 JAPAN
}


\date{\today} 

\begin{abstract}
A new scheme of low energy negative muon source with the muon catalyzed fusion (MuCF) is described.
In the MuCF reaction process, muonic helium ions ($\mu$He$^+$) are created.   By re-accelerating and stripping $\mu$He$^+$ ions, a low emittance negative muon beam is generated.  
\end{abstract}

\maketitle


\section{\label{sec:level1}Introduction}

Negative muons are produced by the decay of negative pions.
The mass is 105.6 MeV, about 200 times that of an electron.
The lifetime is 2.2 $\mu$s.
One of their important characteristics is that negative muons create muonic atoms.
Since the muon mass is 200 times larger than that of an electron, the separation from the nucleus is 200 times smalerl and the binding energy of the negative muon in the muonic atom is 200 times larger.


When negative muons are captured by atoms, high-energy X-rays are produced as they settle down to a stable state, called muon K-X-rays.
In atoms with an atomic number Z of 20 or more, the negative muon is captured by the nucleus.
The captured nucleus is transmuted into another nucleus/isotope by $\beta$-decay and neutron emission.
Utilizing this, nuclear transmutation by shortening the life of long-lived nuclear waste is expected.

Muon-catalyzed nuclear fusion with negative muons is also an interesting application.
When a negative muon is captured by a DT molecule, the distance between DT nuclei approaches 1/200 causing a nuclear fusion reaction.
After the nuclear fusion reaction, the negative muon is again captured by another DT molecule, causing repeated nuclear fusion reactions.
This continues for the lifetime of the negative muon, theoretically enough time for nearly 1000 times.
In reality, the number of reactions is limited to about 150 times because the muon can be adsorbed by the helium nucleus produced by the nuclear fusion reaction.

In high energy particle physics, a muon collider(MuC) has been discussed as a future high energy lepton collider\cite{muc}.
The current global standard scenario of MuC is based on beam cooling with ionization to achieve the luminosity of colliding beams required for the physics goals. 
The emittance reduction at the first stage of the beam cooling in this scenario is about 1/100 and the expected normalized emittance for both transverse and longitudinal directions is about 1.5 $\pi$mm.

In the case of positive muons, it may be possible to obtain low emittance and high brightness beams by a method other than ionization cooling. 
There are a couple of ideas to produce the bright positive muon beams, and some of them are under development.
One is a scheme of dissociating positive muons from ”muonium” by a laser ionization method and extracting it as a positive muon beam. 
Muonium is a particle in which a positive muon traps an electron, just like a hydrogen atom. 
This is under development at KEK-JPARC.\cite{kek}
The other is a friction cooling method that is now under development at PSI.\cite{psi} 
The principle is that positive muons are slowed down in low-temperature helium gas to about 0.2 MeV/c by collisions (friction) with the low-temperature wall and the low-temperature gas. 
At energies below 0.2 MeV/c, positive muons capture electrons and become muonium. 
Therefore, their energy must be cooled to around the energy at which electron exchange occurs. 
In this scheme, a DC beam is possible.

For negative muons, unfortunately, no practical scheme has been proposed that could be applied to the Muon Collider,. 
The frictional cooling scheme cannot be applied to negative muons. 
When $\pi^-$ enters the material, it is absorbed and its decay to $\mu^-$ is greatly suppressed.
Another difficulty is that negative muons are easily captured by atoms at low energy and become muonic atoms. 
Their high dissociation energy ( ionization energy) makes it difficult to dissociate them again and form them as a beam.

We propose a new method for generating negative muon beams with low emittance and high brightness. 
The method is to produce negative muonic He ions ($\mu$He$^+$ ) and accelerate them to an energy sufficient to detach in the materials (gas or thin film), that generate a negative muon beam.
In order to generate energetic muonic ions, a muon-catalyzed fusion that produce energetic $\mu$He$^+$ ions can be used.

\section{Generation of negative muon beam with accelerated $\mu$He$^+$ ions}

If the MuCF target region is small enough to extract and accelerate $\mu$He$^+$ ions before muon stripping, they could be a good source of negative muons.
Based on this idea, we have proposed a new scheme for generation of low emittance low-energy negative muon beam.

A schematic layout of the system is shown in Fig.1.
Negative muons produced by a hadron accelerator are slowed down to about 1 MeV/c or less and introduced into the muon-catalyzed fusion region. 

\begin{figure}
\includegraphics*[width=110mm]{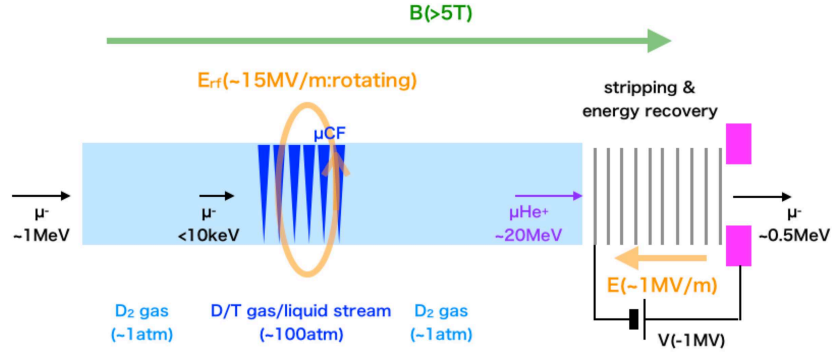}
\centering
\caption{Schematic diagram of negative muon beam generation}
\label{f1}
\end{figure}

Muon-catalyzed fusion reactions were discovered by L.W. Albaretz in 1956 in a hydrogen bubble chamber experiment.\cite{alva}
When a muonic D-T molecule is formed in which the negative muon is captured by a D-T molecule, the D and T nuclei are brought 200 times closer together, and nuclear fusion occurs immediately.
The captured negative muon is released after fusion, and captured again by another D-T molecule.
Thus, the reactions continue.
The fusion catalysis occurs roughly 1500 times during the negative muon lifetime (2.2 $\mu$sec) if there is no disturbance.
However, there is a chance to capture a negative muon by the $\alpha$ particle produced in a D-T fusion reaction.
This occurs about once per about every 100 fusions. 
Once the negative muon is captured by the $\alpha$ particle, the reactions stop and limits the energy production rate.
To overcome this restriction, recently, a new method  was proposed by the author.\cite{mori}

In the muon-catalyzed fusion region, DT fusion reactions are induced in the dense DT molecular medium (gas or liquid at several hundred atmospheres) by catalysis with negative muons.
The negative muons are captured by the helium nuclei created in the fusion reaction and muonic helium ions ($\mu$He$^+$) after being produced. 
Thus, when the $\mu$He$^+$ ions collide with a thin film such as Be and accelerated to an energy sufficient to detach the negative muons, a low energy negative muon beam can be generated.
In this scheme, the initial $\mu$He$^+$ ions is mono-energetic (3.5MeV).
Therefore, the emittance and energy spread of negative muon beam after stripping are expected to be small.

There are two major difficulties to realize this scheme: one is neutralization by electron capture and the other negative muon energy loss through the stripping process.
 




\subsection{Neutralization by electron capture}

As shown in Fig. 2, if the $\mu$He$^+$ ion energy is less than 1 MeV, the ion captures an electron and is neutralized.
The acceleration of $\mu$He$^+$ ions becomes difficult.
In the MuCF reaction, fortunately, the initial energy of the $\mu$He$^+$ ion is 3.5 MeV, which can escape neutralization.
But, while passing through the medium, it is slowing down and captures an electron. 
Thus, energy recovery and acceleration of $\mu$He$^+$ ions are required to avoid neutralization.
In order to do this, the cyclotron resonance acceleration with RF field can be used.

To reduce the energy loss in the DT target as much as possible, the DT target is spatially localized.
The efficiency of energy recovery and re-acceleration under these conditions can be enhanced by a spatially localized DT target.\cite{mori} 

\begin{figure}
\includegraphics*[width=70mm]{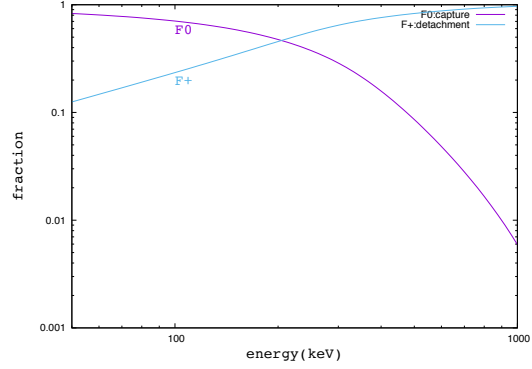}
\centering
\caption{Charge equilibrium distribution of $\mu$He$^+$ and $\mu$He$^0$}
\end{figure}

\subsection {Energy loss of negative muon through stripping}

Another issue is an energy loss of $\mu^-$ through stripping.
A negative muon beam can be generated from a $\mu$He$^+$ ion beam by a stripping reaction.

\begin{eqnarray}
\mu He^+ + X \rightarrow \mu^- + He^{2+}+ X.
\end{eqnarray} 

Immediately after stripping, the $\mu$He$^+$ ion velocity and the $\mu^-$ velocity are almost equal.
Therefore, the energy losses of $\Delta E$ of the $\mu$He$^+$ ion and negative muon in the material are almost the same.
To extract the negative muon, the negative muon energy must be greater than the energy loss.

In other words, $E_{\mu}$ must be larger than $\Delta E$.
The probability of $\mu$He$^+$ ions being stripped into negative muons and helium ions is given by the stripping cross section $\sigma(E)$, stopping power  $S(E)$ and material molecular density n as follows.

\begin{eqnarray}
\alpha=\exp\left[-\int_{E_0}^{E_f}\frac{\sigma(E)n}{S(E)}dE\right].
\end{eqnarray}
Here, $E_0$ and $E_f$ are the initial and final energies of $\mu$He$^+$ ion, respectively. 

The energy loss $\Delta E=E_0 - E_f$, where this is 1/e, is given by,

\begin{eqnarray}
\Delta E \sim -\frac{S(E_0)}{\sigma(E_0)n}.
\end{eqnarray}

From the negative muon energy $E_{\mu} > \Delta E$, we have,

\begin{eqnarray}
E_f>\left[-\frac{S(E_0)}{\sigma(E_0)n}\right]\frac{m_{\mu He}}{m_{\mu}}.
\end{eqnarray}

Here is an example.
When the stripping target is a Be foil with a thickness of 4.8 $\mu$m, $\Delta E$ and $E_f$ are given from Eqs.(3) and (4) as,  

\begin{eqnarray}
\Delta E= 2 MeV \  and  \ E_f > 80 MeV.
\end{eqnarray}
respectively.
Thus,

\begin{eqnarray}
E_0 > E_f+\Delta E=82 MeV.
\end{eqnarray}
Accelerating to this energy requires an accelerator with protons equivalent to several hundred MeV.
If $S(E_0)$ can be reduced, $E_f$ becomes small.
Effective reduction of $S(E_0)$ can be achieved by energy recovery of stripped negative muons.

\subsection{Energy recovery of negative muon} 

The problem is that the energy loss of the stripped negative muons in the thin film is large.
In order to lower the energy of $\mu$He$^+$ ions, the stopping power $S(E)$ of negative muons at the stripping foil must be effectively decreased.
For that purpose, the energy recovery of negative muon is useful.
If $S(E)$ of the negative muon can be made virtually zero in the stripping process, $E_0$ can be decreased.
This is also the ionization cooling of negative muons.


The energy recovery process of the negative muon beam during stripping is actually ionization cooling itself.
We have evaluated the change in muon beam emittance with the ionization cooling rate equations.
Here, we assume a muon energy of 0.5 MeV.  This corresponds to being stripped from $\mu$He$^+$ ions at 20 MeV. 
The stripping foil is Be and its thickness is 4.8 $\mu m$ for each of the 20 steps of energy recovery. 

The calculation results are shown in Fig.3.
The transverse normalized emittance after 20 steps is 0.2 mm., and the energy spread caused by straggling is about 150  keV. 

\begin{figure}
\includegraphics*[width=70mm]{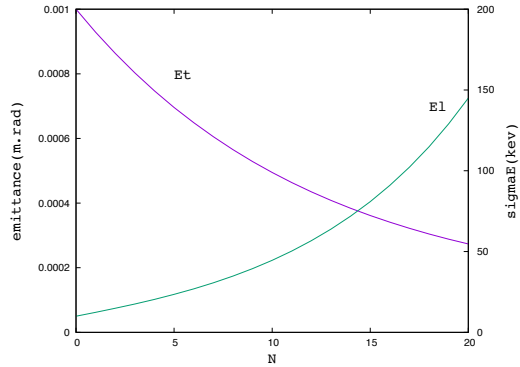}
\centering
\caption{Emittance behaviors of the negative muon beam through the ionization cooling process.}
\label{f1}
\end{figure}

\section{Characteristics of the negative muon beam}

A global simulation was performed to evaluate the characteristics of the negative muon beam by this scheme.

\begin{figure}
\includegraphics*[width=70mm]{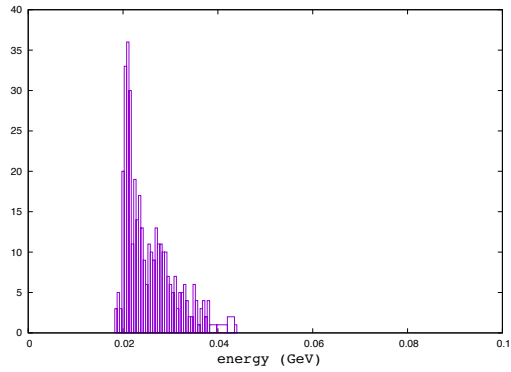}
\centering
\caption{Simulation results of $\mu$He$^+$ ion energy distribution}
\end{figure}

Low energy negative muons produced by a hadron accelerator are introduced into the MuCF region.
$\mu$He$^+$ ions are formed in the MuCF reaction and accelerated by an RF field associated with cyclotron resonance.
They are stripped to negative muons by Be thin foils at the energy recovery and ionization cooling regions.
The energy recovery and ionization cooling region consist of about 20 thin Be foils and an electrostatic field of about 1-2 MV/m.


Figure 4 shows the simulation results of $\mu$He$^+$ ion energy distribution at the entrance of the stripping region.
Here, we assumed the RF field of 30 MV/m for acceleration.
The total number of $\mu$He$^+$ ions is about 30 \% of the initial number of  $\mu$He$^+$ ions generated in the MuCF region.
Using this energy distribution, we have evaluated the characteristics and performance of the negative muon beam  generated through the processes of stripping, energy recovery, and ionization cooling.  

Figure 5 shows the results of the tracking simulation on the momentum and beam spread distribution of the negative muon beam. 
Characteristics of the negative muon beam are examined with simulations using the G4BL tracking code.  As initial particle distributions in 3D directions, gaussian distributions are assumed where each standard deviation $\sigma$ is adopted from the simulation results of $\mu$He$^+$ ions. 
In the longitudinal direction, the momentum spread was evaluated as,

\begin{eqnarray}
\sigma_l=\sigma_{\mu He} + \frac{EL}{\sqrt{2}},
\end{eqnarray}
where $\sigma_{\mu He}$ is the momentum spread of the $\mu$He$^+$ ion beam, $E$ is the electric field strength and $L$ is the length of the energy recovering system.

The emittance of negative muon beam is evaluated from these tracking simulations.
The transverse emittance is 1.5 $\pi$mm and the longitudinal emittance is 2.5 $\pi$mm.
These values are almost the same or even smaller than those obtained from ionization cooling using the MAP strategy.\cite{muc}

\begin{figure}
\includegraphics*[width=40mm]{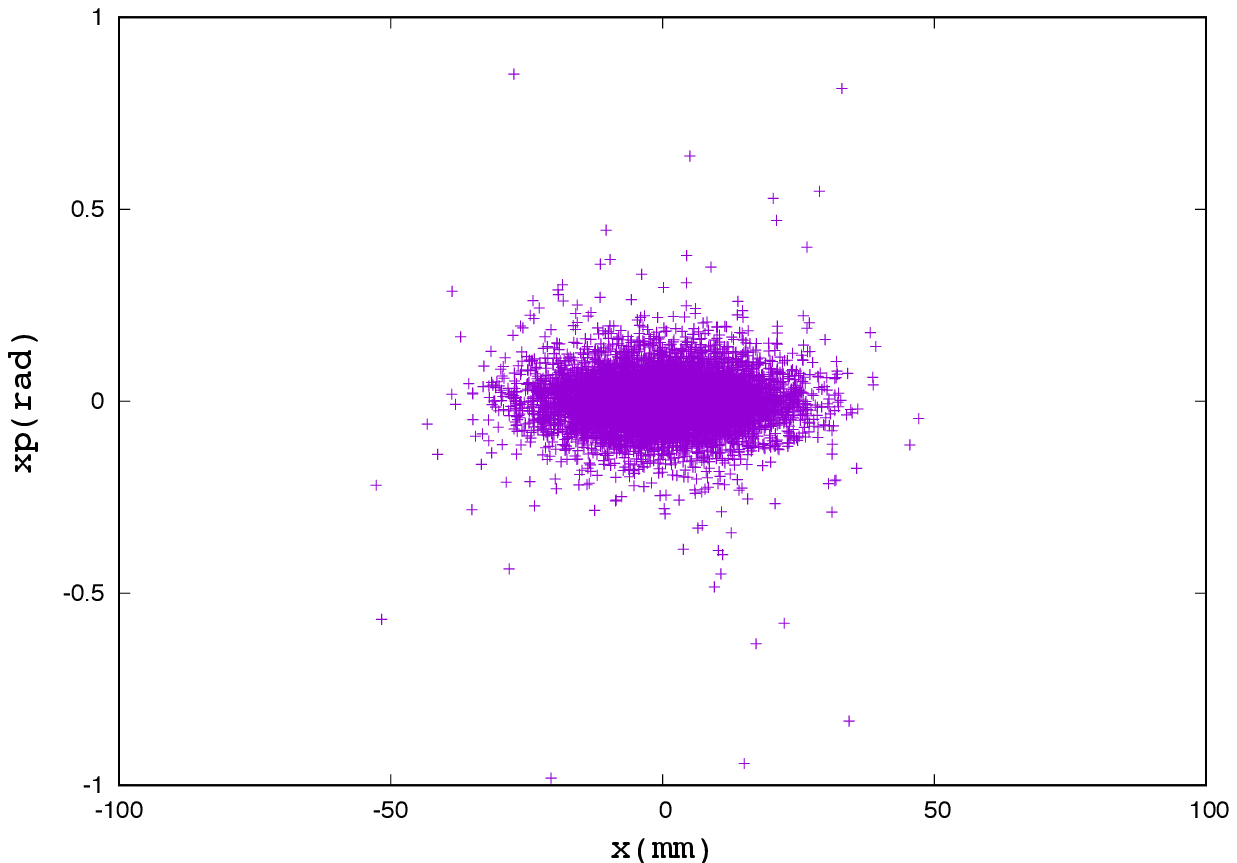}
\includegraphics*[width=40mm]{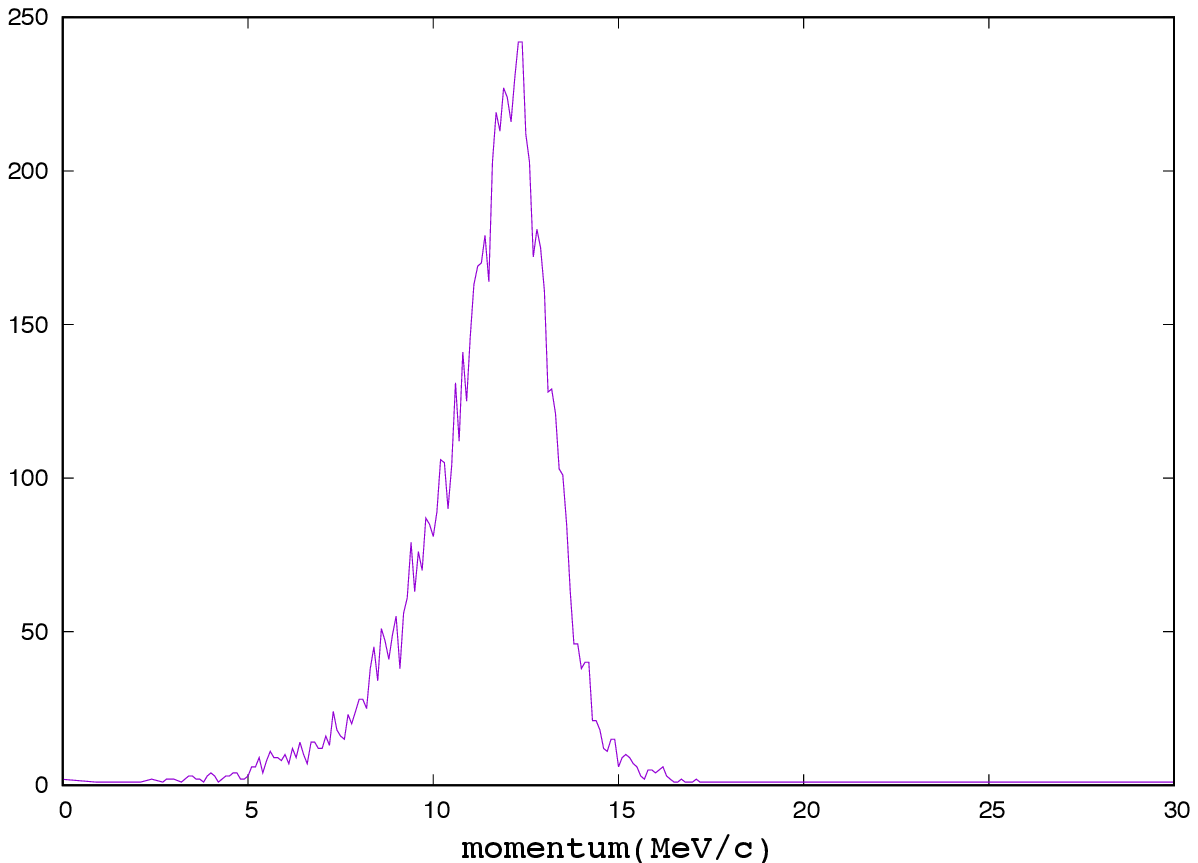}
\includegraphics*[width=35mm]{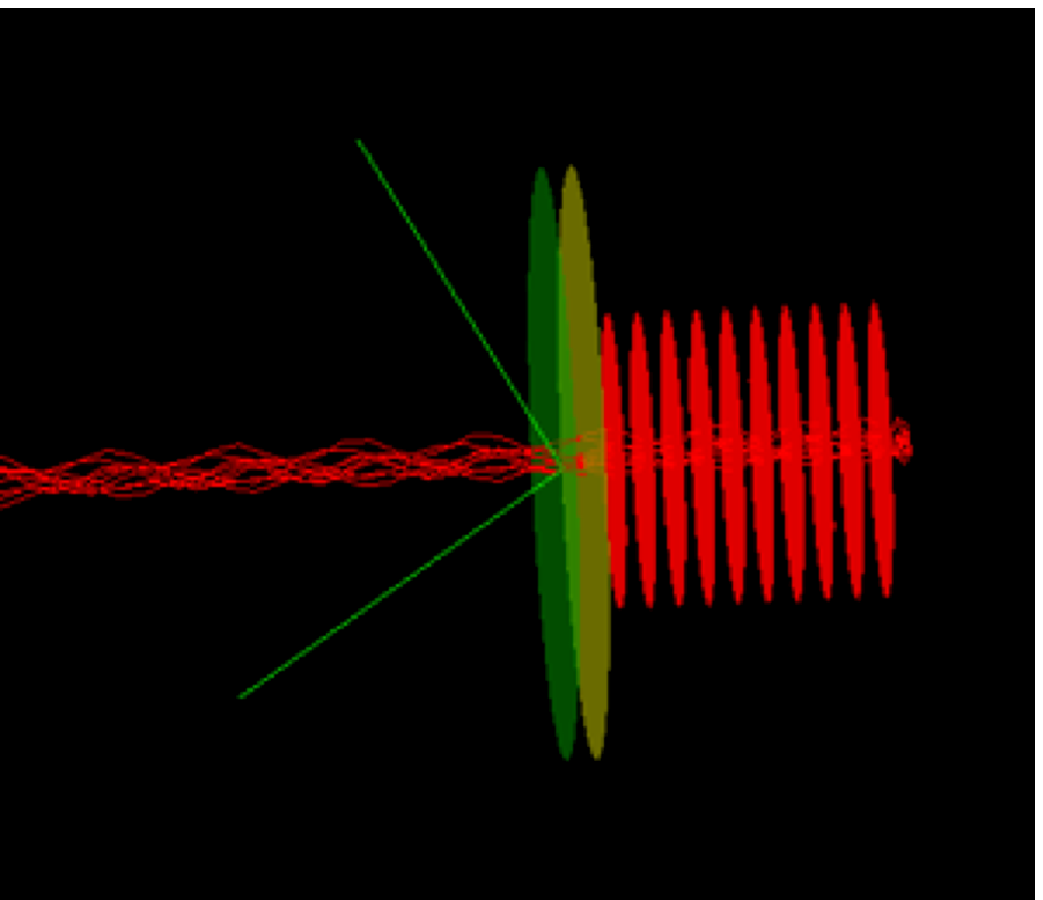}
\centering
\caption{Simulation results of the momentum and beam spread distributions of the negative muon beam.(Left:emittance Middle:momentum Right:G4BL-Visualout)}
\end{figure}

\section{Summary}

A new method of negative muon beam generation that efficiently accelerates $\mu$He$^+$ ions produced in the MuCF reaction and strips negative muons from them was described.

This method should solve two problems:

(1)To avoid electron capture,  $\mu$He$^+$ ions must be accelerated from an initial energy of 3.5 MeV to about 20 MeV without decelerating.

(2)Energy recovering is necessary to reduce energy loss in the negative muon stripping region.

Overcoming the above two points is the most important key.
Simulation calculations show that, as beam emittance,

\begin{eqnarray}
\epsilon_t = 1.5 \pi mm, 
\epsilon_l = 2.5 \pi mm.
\end{eqnarray}
These values are comparable to or smaller than the beam emittance after ionization cooling at the muon collider.

The technical challenge in this scheme is to accelerate the $\mu$He$^+$ ions created in the MuCF region to a given energy. In other words, the $\mu$He$^+$ ions must be accelerated from 2 MeV (0.5 MeV/u) or higher to about 20 MeV, which is an energy that is not neutralized by electron capture. 
For that purpose, it is necessary to efficiently generate the rotating (circularly polarized) RF electric field (E=15 MV/m, f=76 MHz).

If this method is realized, it will be possible to accelerate continuous negative muon beams to high energies with conventional hadron accelerators (cyclotron, linac, etc.).
Various applications of mono-energetic high-energy negative muon beams of several 100 MeV to several GeV becomes possible in the scientific field.

\begin{acknowledgments}
The author would like to express his sincere appreciations to Drs. C.D.P. Levy and Y. Ishi for their helpful and valuable comments on this work.
\end{acknowledgments}

\nocite{*}
\bibliography{aipsamp}

\appendix

\section{Sticking probability of negative muons as a function of RF field strength}

The simulation results of the negative muon sticking probability as a function of the RF electric field strength are shown in Fig.A.1

In achieving break-even muon-catalyzed fusion, a RF field strength of 15 MV/m is sufficient to achieve physical break-even in a linearly polarized electric field. 
Engineering break-even can also be achieved if a circularly polarized electric field is used.
Two cases of electric fields are assumed here: a linearly polarized electric field and a circularly polarized(rotating) electric field.

As can be seen, the circularly deflected electric field is more efficient than the linearly deflected electric field.
This is due to the fact that the acceleration efficiency, i.e. the transit time coefficient, is almost one in the case of the circularly deflected electric field.

\renewcommand{\thefigure}{A.\arabic{figure} }
\setcounter{figure}{0}

\section{Generation of $\mu$He$^+$ ion in He gas}

\renewcommand{\theequation}{A.\arabic{equation} }
\setcounter{equation}{0}

In helium gas of several 10 atms, negative muonic helium ions ($\mu He^+$) can be easily formed because the biding energy of a negative muon is 200 times larger than an electron.\cite{rosen}\cite{krau}
However, it is difficult to extract and accelerate them to form $\mu$He$^+$ ion beams.

A $\mu$He$^+$ ion behaves just like a proton from the Coulomb force point of view.
Therefore, as a $\mu$He$^+$ ion is accelerated by the electric field, it can capture an electron and form a neutral  
$\mu$He atom.
In order to form the $\mu$He$^+$ ion beam, the initial energy of$\mu$He$^+$ ion produced in He gas must be  larger than 1MeV.   Otherwise, $\mu$He$^+$ ions capture the electrons to form the neutral $\mu$He atoms immediately.
To accelerate $\mu$He$^+$ ions up to more than 1MeV before neutralization by electron capture, at least an electric field strength of more than 1GV/m is necessary.
Thus, it is impossible to form the $\mu^-$ beam with this scheme!

\begin{figure}
\includegraphics*[width=100mm]{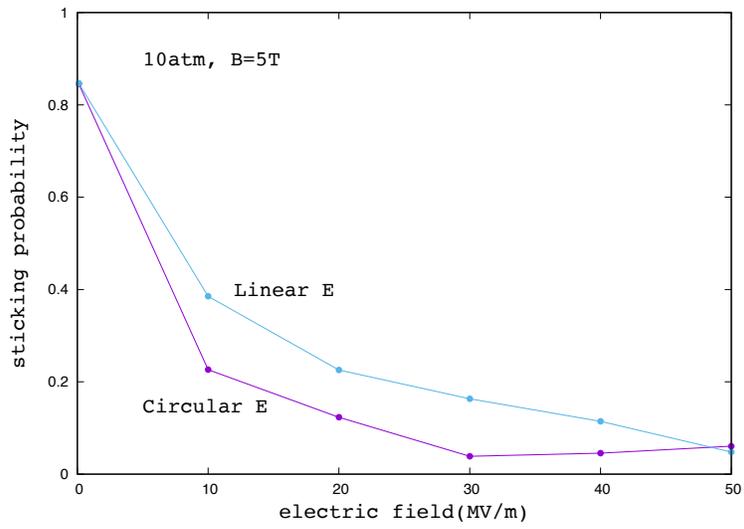}
\centering
\caption{Simulation results on the momentum and beam spread distribution of the negative muon beam.}
\end{figure}

\end{document}